# Integrated photonic platform for quantum information with continuous variables

Francesco Lenzini[1,2], Jiri Janousek[3,4], Oliver Thearle[3,4], Matteo Villa[1], Ben Haylock[1], Sachin Kasture[1], Liang Cui[5], Hoang-Phuong Phan[6,7], Dzung Viet Dao[6,7], Hidehiro Yonezawa[8], Ping Koy Lam[4], Elanor H. Huntington[3], and Mirko Lobino[1,6,*]

[1] *Centre for Quantum Computation & Communication Technology and Centre for Quantum Dynamics, Griffith University, Brisbane QLD 4111, Australia*
[2] *Institut of Physics, University of Muenster, 48149 Muenster, Germany*
[3] *Centre for Quantum Computation & Communication Technology and Research School of Engineering, The Australian National University, Canberra ACT 2601, Australia*
[4] *Centre for Quantum Computation & Communication Technology and Department of Quantum Science, Research School of Physics and Engineering, The Australian National University, Canberra ACT 2601, Australia*
[5] *College of Precision Instrument and Opto-electronics Engineering, Tianjin University, Tianjin 300072, China*
[6] *Queensland Micro- and Nanotechnology Centre, Griffith University, Brisbane QLD 4111, Australia*
[7] *School of Engineering, Griffith University, Brisbane QLD 4111, Australia*
[8] *Centre for Quantum Computation & Communication Technology and School of Engineering and Information Technology, The University of New South Wales, Canberra, ACT 2600, Australia*

*\* m.lobino@griffith.edu.au*

Integrated quantum photonics provides a scalable platform for the generation, manipulation, and detection of optical quantum states by confining light inside miniaturized waveguide circuits. Here we show the generation, manipulation, and interferometric stage of homodyne detection of non-classical light on a single device, a key step towards a fully integrated approach to quantum information with continuous variables. We use a dynamically reconfigurable lithium niobate waveguide network to generate and characterize squeezed vacuum and two-mode entangled states, key resources for several quantum communication and computing protocols. We measure a squeezing level of $-1.38 \pm 0.04$ dB and demonstrate entanglement by verifying an inseparability criterion $I = 0.77 \pm 0.02 < 1$. Our platform can implement all the processes required for optical quantum technology and its high nonlinearity and fast reconfigurability makes it ideal for the realization of quantum computation with time encoded continuous variable cluster states.

Integrated quantum photonics [1] has emerged as the ideal platform for the implementation of optical quantum computation [2], communication [3], and sensing protocols [4]. By confining light inside miniaturized waveguide circuits, it is possible to generate quantum states of light [5,6], interfere them over waveguide networks [7], and detect them with integrated detectors [8]. Integration of these three key operations establishes the stability and scalability of this technology, enabling a continuous increases in complexity and capabilities of these devices [9,10].

Optical quantum information is more commonly encoded in one of the discrete degrees of freedom (or discrete variables, DV) of single photons such as polarization [11] or path [2]. This approach enables operations with near-unity gate fidelity [12], but is currently limited by the lack of on-demand single photon sources and deterministic two photon quantum gates. The encoding of information on operators that are continuous variables (CV) such as quadrature amplitudes, offers the advantages of deterministic generation of quantum states and operation, at the expense of a higher tendency to imperfect gate fidelities [13]. This approach has been demonstrated in several fields, including secure quantum communication [14], quantum enhanced sensing [15], and quantum information processing [16]. Hybrid approaches combining the benefits of DV and CV systems have also been proposed and experimentally demonstrated [17].

While integrated optics provides great stability and scalability to all types of encoding, in CV schemes it also greatly simplifies the configuration of current experimental setups, replacing phase-locked cavities for generation of squeezed light with single-pass waveguides [18]. It also eliminates the need of mode-cleaning cavities for homodyne detection thanks to the nearly-perfect overlap between optical modes in guiding structures [19,20].

Furthermore, the possibility of achieving broad generation bandwidths in a single-pass squeezer [21], and performing fast-switching operations with electro-optically tunable waveguides [11,22], makes integrated optics an attractive platform for the implementation of frequency [23,24] or time-multiplexed encoding [25,26] and fast feedforward operations needed for CV measurement-based quantum computing [27].

Here we demonstrate a nonlinear and reconfigurable integrated device which generates, actively manipulates, and performs the interferometric stage of homodyne detection on non-classical light fields. The device is formed by two integrated sources of squeezed vacuum and electro-optically tunable phase shifters and beam splitters where squeezed vacua can interfere and be characterized. Complemented with photon-number resolving detectors or non-Gaussian ancilla inputs [13], such architecture can enable non-Gaussian operations for universal quantum information processing or hybridisation with DV systems [17].

Figure 1 shows a schematic of the integrated chip and the experimental setup. The device is made of a network of six waveguides patterned on a Z-cut lithium niobate substrate by reverse proton exchange [28] (see Methods for details on the fabrication process).

Two periodically poled waveguides, phase-matched around 1550 nm, are used to generate two squeezed vacuum states which are interfered on a reconfigurable directional coupler (DC1) for the generation of a two-mode CV entangled state [13]. Both waveguides have a 2 cm interaction length, extrapolated from the 0.5 nm FWHM of the second harmonic generation efficiency as a function of the pump wavelength (Fig. 2a). This interaction length corresponds to a 96 nm FWHM bandwidth for the generated squeezed light.

Two directional couplers (DC2, DC3) designed with a 100% splitting ratio (SR) at 1550 nm separate the generated quantum states from the pump beams, which remain confined in the initial waveguides due to the smaller mode field diameter. Balanced homodyne detection is performed by mixing the generated signals with two local oscillator beams in two tunable directional couplers (DC4, DC5). Electrodes patterned on top of the waveguides are used to scan the phase of the local oscillators, and to tune the SR of the directional couplers [22]. Local oscillator phases $\phi_{LO1}$ and $\phi_{LO2}$ are scanned by $2\pi$ when a ±10 V waveform is applied (see Fig. 2b and c), while the splitting ratios of the reconfigurable couplers can be reduced from their no voltage values down to ~0.5% with an applied voltage in the ±20 V range (see Methods). The SRs of DC4 and DC5 are tuned at around 50% for balanced homodyne detection.

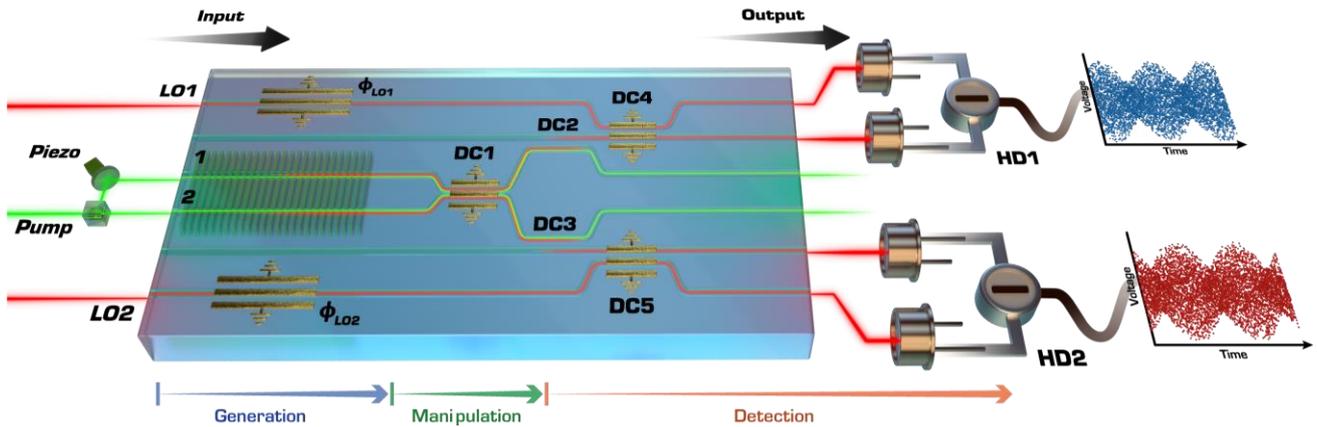

**FIG. 1. Configuration of the chip and experimental setup.** Periodically poled waveguides (1 and 2) are the nonclassical light sources. The squeezed vacuum states are manipulated in a reconfigurable directional coupler (DC1) for the generation of two separable squeezed states or a two-mode CV entangled state. DC2 and DC3 are used to separate the pump at ~777 nm from the signal at ~1554 nm. The rest of the network ($\phi_{LO1}$, $\phi_{LO2}$, DC4 and DC5) is made of the reconfigurable phase shifters and couplers of the two homodyne detectors.

Two mode CV entanglement is generated by interfering orthogonal squeezed vacuum states from the two periodically poled waveguides on DC1 with a SR tuned at 50%. Separable squeezed vacuum states are created when the SR of DC1 is set as close as possible to zero (~0.5%), for implementing the identity operation. Due to imperfections in the waveguide fabrication process the SR of DC2, DC3 was found equal to 80% and 86% respectively, reducing, in this way, the maximum amount of measurable squeezing. In future implementations the performance of the filters could be improved by patterning electrodes with alternating phase mismatch [29], to allow tuning the SR in the full 0-100% range.

The master laser is an amplified cavity diode laser, based on a gain chip [30], and tunable in the 1550 nm wavelength range. The pump beam is obtained by frequency doubling part of the master laser power with a PPKTP crystal in a single-resonant cavity (see Methods), with the remaining power used as local oscillators. All the beams are coupled into the chip with a fibre V-groove array, while the output modes are collected by a 0.5NA lens and sent to a pair of homodyne detectors (HD1, HD2) with 99% quantum efficiencies (see Methods). Electronic filtering is used to select a wide side-band from 4 MHz to 35 MHz, which is measured in the time-domain with a digital oscilloscope. For generation of CV entanglement, the relative phase of the two pump beams is controlled in free-space with a piezo-electric mirror. The second and fifth inputs of the device were unused in this experiment, but in future may be used to implement displacement operations [13,17].

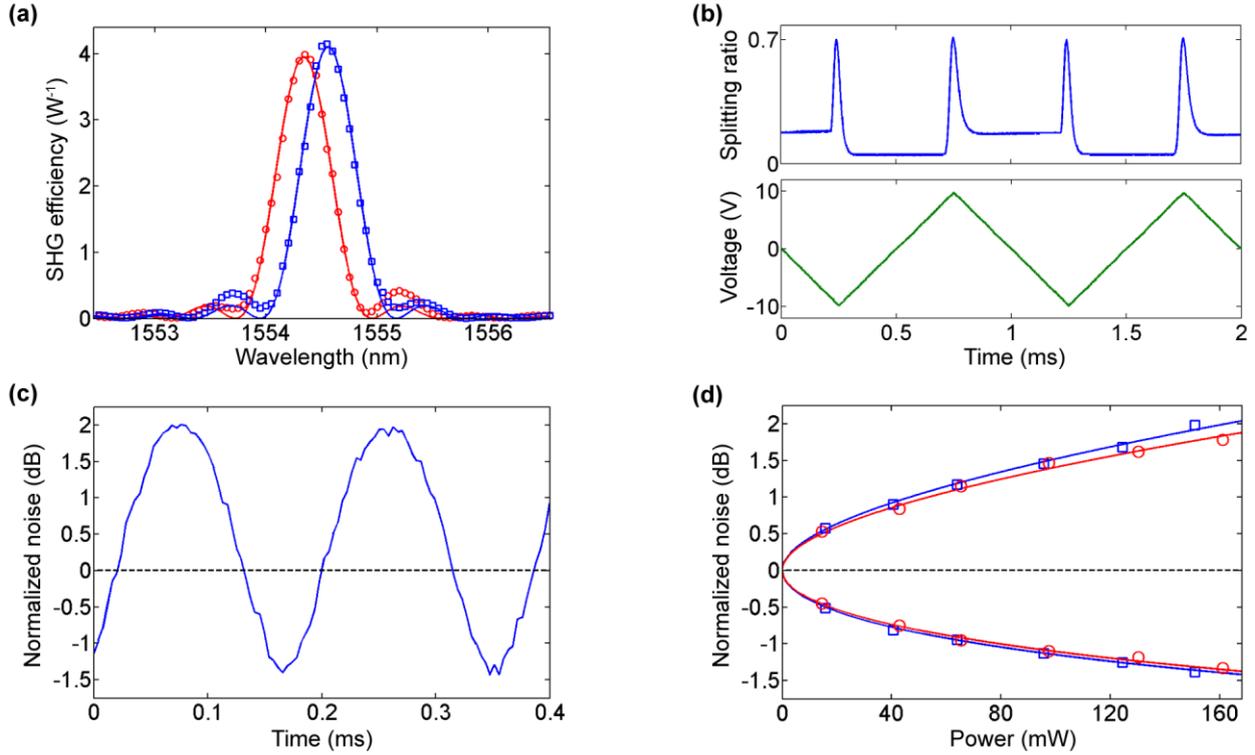

**FIG. 2. Generation and homodyne detection of squeezed vacuum states.** (**a**) Measured normalized second harmonic generation (SHG) efficiencies $P_{2\omega}/P_\omega^2$ for waveguide 1 (blue points) and waveguide 2 (red points) at $T = 125$ °C and relative theoretical fit (solid lines). Pump and second harmonic powers are corrected for Fresnel losses at the output facet and propagation losses (see Methods). The FWHM of the fit sinc² are $\simeq 0.5$ nm, consistent with a $\simeq 2$ cm interaction length. The waveguides have the same normalized conversion efficiency $\simeq 370$ %W⁻¹ at $\lambda = 1554.45$ nm. (**b**), Splitting ratio of DC1 (top image), and voltage applied to the local oscillator phase shifters (bottom image) as a function of time . SR measurement is performed by injecting a 1550 nm beam into waveguide 2 and measuring the power at the first output with a photodiode. DC1 electrode is driven with a square wave with 1 kHz frequency and ±16 V amplitude. Distortion of the square signal is due to the limited bandwidth of the voltage amplifier. (**c**), Noise trace measured from HD1 for a 154 mW pump power. Noise variance is calculated on time intervals of 4 μs duration and averaged over 40 sequential traces. Sampling rate is 50MSPS. (**d**), Measured squeezing and anti-squeezing levels as a function of pump power for waveguide 1 (blue squares) and waveguide 2 (red circles). Solid lines are the fits made with the function of Eq. 1. Error on the measured noise levels is ±0.04 dB. Pump powers are measured at the output of the device and corrected for Fresnel losses at the output facet and propagation losses inside the waveguides. Pump wavelength is $\lambda_P = 1554.55/2$ nm for waveguide 1, and $\lambda_P = 1554.35/2$ nm for waveguide 2.

The device was first configured for generation and homodyne detection of squeezed vacuum states. SR of DC1 was set to the minimum value of 0.5%, and the phases of local oscillator beams scanned by approximately 2π with a ramp function (see Fig. 2b). To prevent accumulation of surface charge due to the application of a DC offset, the three directional couplers were modulated by a square function with zero mean amplitude and a 1 kHz frequency. Post-processing on the acquired signal was used to select a 0.4 ms time window centred around the applied square waves every modulation period. Characterization was carried out by injecting a pump beam into each periodically poled waveguide at a time, and measuring the resulting noise levels on the adjacent homodyne detectors. Due to the high operational temperature ($T = 125$ °C), the coupled pump power inside the waveguides can be increased up to $\simeq 160$ mW without any evidence of photorefractive damage.

Figure 2c shows the noise trace from HD1, corresponding to a maximum measured squeezing (anti-squeezing) level $\langle\Delta^2 \hat{X}^-\rangle = -1.38 \pm 0.04$ dB ($\langle\Delta^2 \hat{X}^+\rangle = 1.98 \pm 0.04$ dB) for a pump power $P = 154$ mW. After correcting for 13% Fresnel losses, which could be eliminated with an anti-reflection coating on the output facet, and inefficiencies of the filter (SR=80%), we estimate that -2.15±0.04 dB of squeezing is generated in our device. The squeezing and anti-squeezing levels measured for both waveguides as a function of pump power are shown in Fig. 2d. The points are fitted using the function

$$\langle \Delta^2 \hat{X}^{\pm} \rangle = \eta e^{\pm 2\mu \sqrt{P}} + 1, \qquad (1)$$

where $\eta$ is the overall detection efficiency. Results of the fit give $\mu_1 = 0.030 \pm 0.001$ mW$^{-1/2}$, $\mu_2 = 0.027 \pm 0.001$ mW$^{-1/2}$, $\eta_1 = 0.52 \pm 0.02$, and $\eta_2 = 0.54 \pm 0.02$, against estimated $\eta_1 = 0.55$, and $\eta_2 = 0.6$, for 0.14 dB/cm propagation losses (see Methods). We note that $\eta_1$ is found compatible, within the 95% confidence interval, with the estimated value. For waveguide 2, the extra 0.06 inefficiency is likely introduced by imperfections in the waveguides along the path of the generated signals.

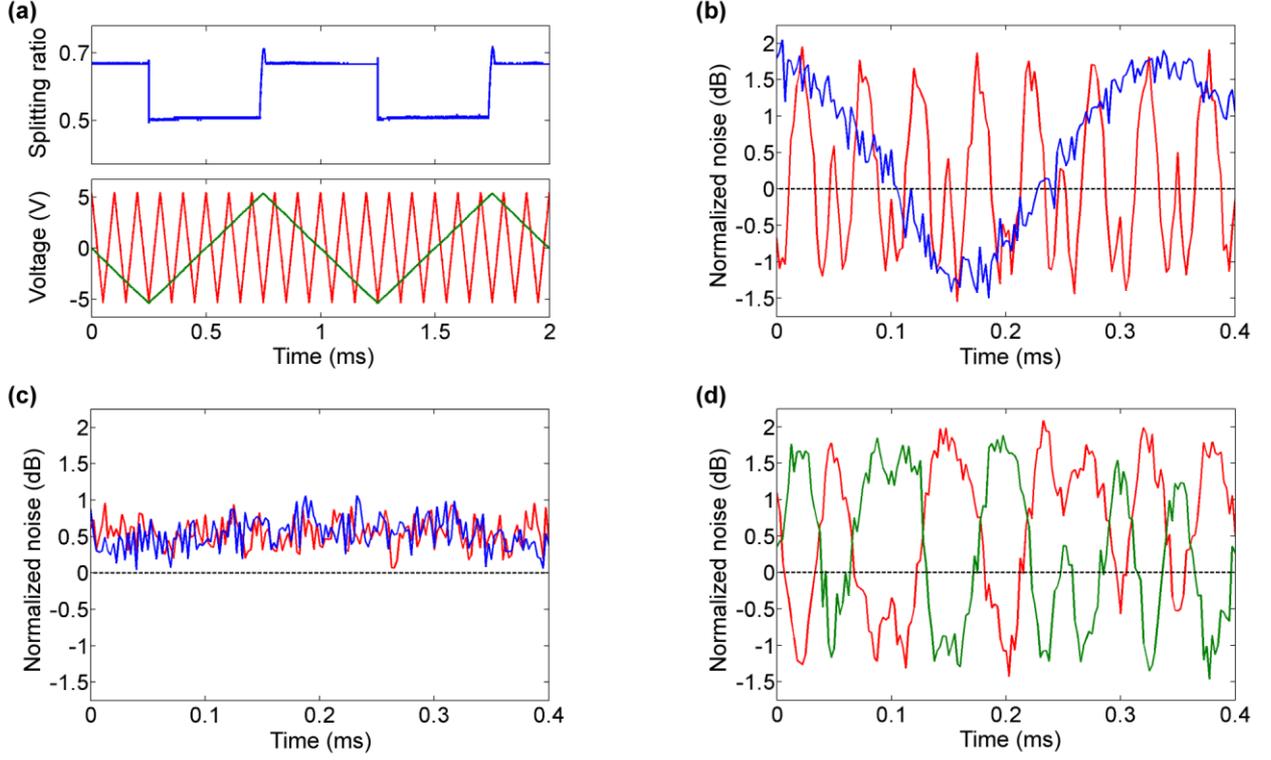

**FIG. 3**. **Generation and characterization of CV entanglement.** (**a**) Splitting ratio of DC1 (top image), and voltage applied to phase shifters $\phi_{LO1}$ (green trace, bottom image) and $\phi_{LO2}$ (red trace, bottom image) as a function of time. SR measurement is performed by injecting a 1550 nm beam into waveguide 2 and measuring the power at the first output with a photodiode. DC1 electrode is driven with a square wave at 1 kHz frequency and ±5.5 V amplitude. Scanning frequency is 1 kHz for $\phi_{LO1}$ and 10 kHz for $\phi_{LO2}$. (**b**) Noise levels measured from HD1 (blue trace) and HD2 (red trace) when the pump beams are in phase. (**c**) Noise levels measured from HD1 (blue trace) and HD2 (red trace) when the pump beams are out of phase. (**d**) Noise levels of summed quadratures $\langle \Delta^2(\hat{X}_1 + \hat{X}_2) \rangle$ (green trace) and subtracted quadratures $\langle \Delta^2(\hat{X}_1 - \hat{X}_2) \rangle$ (red trace) when the pump beams are out of phase. Noise variance is calculated on time intervals of 2.5 µs duration and averaged over 10 sequential traces. Sampling time is 20 ns. Measurements are performed with two pump beams with $P = 122$ mW and $\lambda_P = 1554.45/2$ nm.

Next, the device was configured for the generation and characterization of CV entanglement between the two spatial modes after DC1. SR of DC1 was set to 50% and the phases of the two local oscillator beams were scanned by approximately $\pi$ at 1 KHz for $\phi_{LO1}$ and 10 KHz for $\phi_{LO2}$ (see Fig. 3a). The phase of pump 1 was scanned simultaneously by approximately $2\pi$ at a much lower speed (50 Hz) using a piezo-electric mirror (see Fig. 1). Entanglement was verified using the inseparability criterion for Gaussian states [31]

$$I = \sqrt{\min[\langle \Delta^2(\hat{X}_1^+ \pm \hat{X}_2^+) \rangle] \times \min[\langle \Delta^2(\hat{X}_1^- \pm \hat{X}_2^-) \rangle]} < 1, \qquad (2)$$

where we use the product form of Ref. 31, and $X^-, X^+$ are, respectively, squeezed and anti-squeezed quadratures when the pump beams are in phase.

Figures 3b-d show the results of the measurements for a 122 mW pump power coupled in each waveguide. When the pump beams are in phase (Fig. 3b), the device generates two separable squeezed states with similar squeezing and anti-squeezing levels, $\langle \Delta^2 \hat{X}_1^- \rangle = -1.16 \pm 0.06$ dB ($\langle \Delta^2 \hat{X}_1^+ \rangle = 1.71 \pm 0.06$ dB) for HD1, and $\langle \Delta^2 \hat{X}_2^- \rangle = -1.11 \pm 0.06$ dB ($\langle \Delta^2 \hat{X}_2^+ \rangle = 1.65 \pm 0.06$ dB) for HD2. When the pump beams have a $\pi$ relative phase (Fig. 3c), as expected

for an entangled state, we observed phase independent and constant noise levels $\langle\Delta^2\hat{X}_1\rangle = 0.53 \pm 0.20$ dB for HD1, and $\langle\Delta^2\hat{X}_2\rangle = 0.54 \pm 0.20$ dB for HD2. Conversely, variance of summed and subtracted quadratures (green and red traces in Fig. 3d) show a phase-sensitive behaviour with correlations below the equivalent shot noise level resulting from the combination of the two homodyne currents (see Methods). From the data of Fig. 3d we calculated $\min[\langle\Delta^2(\hat{X}_1^+ \pm \hat{X}_2^+)\rangle] = -1.19 \pm 0.12$ dB, and $\min[\langle\Delta^2(\hat{X}_1^- \pm \hat{X}_2^-)\rangle] = -1.07 \pm 0.12$ dB, corresponding to $I = 0.77 \pm 0.02 < 1$ which satisfies the inseparability criterion by 11 standard errors.

In conclusion we demonstrated the generation, manipulation, and characterization of nonclassical quantum states of light in a monolithically integrated device. We have shown the reconfigurability of our technology by generating squeezed vacua and CV quadrature entanglement in two separate spatial modes. We calculated that by increasing the peak power to 500 mW with a pulsed pump laser and the interaction length to 4 cm, our fabrication technology could reach ~7 dB of squeezing. Furthermore, the recently developed low loss, high confinement ridge waveguides in lithium niobate [32] can potentially generate more than 10 dB of squeezing with this same material and provide a technology with a footprint similar to the silicon-on-insulator platform. In the future it will be of interest to measure these broadband quantum states with high bandwidth homodyne detectors and demonstrate fast-switching operations in the GHz regime for measurement based quantum computation.

## ACKNOWLEDGEMENTS

We thank Nicholas Robins for lending us the laser. This work was supported by the Australian Research Council (ARC) Centre of Excellence for Quantum Computation and Communication Technology (CE170100012), and the Griffith University Research Infrastructure Program. BH and MV are supported by the Australian Government Research Training Program Scholarship. This work was performed in part at the Queensland node of the Australian National Fabrication Facility, a company established under the National Collaborative Research Infrastructure Strategy to provide nano- and microfabrication facilities for Australia's researchers.

## AUTHOR CONTRIBUTIONS

F.L., J.J., and C.L. performed the experimental measurements. F.L., S.K., B.H., and M.V. designed and fabricated the integrated device. J.J. and O.T. designed and built the SHG cavity and homodyne detectors. H.-P.P. performed electrode wire bonding. D.V.D., H.H, P.K.L., E.H. and M.L. supervised the project. F.L. and M.L. wrote the manuscript with contributions from all authors.

## REFERENCES

[1] S. Tanzilli, A. Martin, F. Kaiser, M. P. de Micheli, O. Alibart, and D. B. Ostrowsky, Laser Photonics Rev. **6**, 115 (2012).

[2] A. Politi, J. C. F. Matthews, and J. L. O'Brien, Science **325**, 1221 (2009).

[3] P. Zhang, K. Aungskunsiri, E. Martín-López, J. Wabnig, M. Lobino, R. W. Nock, J. Munns, D. Bonneau, P. Jiang, H. W. Li, A. Laing, J. G. Rarity, A. O. Niskanen, M. G. Thompson, and J. L. O'Brien, Phys. Rev. Lett. **112**, 130501 (2014).

[4] A. Crespi, M. Lobino, J. C. F. Matthews, A. Politi, C. R. Neal, R. Ramponi, R. Osellame, and J. L. O'Brien, Appl. Phys. Lett. **100**, 233704 (2012).

[5] M. E. Anderson, J. D. Bierlein, M. Beck, and M. G. Raymer, Opt. Lett. **20**, 620 (1995).

[6] H. Jin, F. M. Liu, P. Xu, J. L. Xia, M. L. Zhong, Y. Yuan, J. W. Zhou, Y. X. Gong, W. Wang, and S. N. Zhu, Phys. Rev. Lett. **113**, 103601 (2014).

[7] J. W. Silverstone, D. Bonneau, K. Ohira, N. Suzuki, H. Yoshida, N. Iizuka, M. Ezaki, C. M. Natarajan, M. G. Tanner, R. H. Hadfield, V. Zwiller, G. D. Marshall, J. G. Rarity, J. L. O'Brien, and M. G. Thompson, Nat. Photonics **8**, 104 (2014).

[8] W. H. P. Pernice, C. Schuck, O. Minaeva, M. Li, G. N. Goltsman, A. V. Sergienko, and H. X. Tang, Nat. Commun. **3**, 1325 (2012).

[9] J. Carolan, C. Harrold, C. Sparrow, E. Martin-Lopez, N. J. Russell, J. W. Silverstone, P. J. Shadbolt, N. Matsuda, M. Oguma, M. Itoh, G. D. Marshall, M. G. Thompson, J. C. F. Matthews, T. Hashimoto, J. L. O'Brien, and A. Laing, Science **349**, 711 (2015).

[10] J. Wang, S. Paesani, Y. Ding, R. Santagati, P. Skrzypczyk, A. Salavrakos, J. Tura, R. Augusiak, L. Mančinska, D. Bacco, D. Bonneau, J. W. Silverstone, Q. Gong, A. Acín, K. Rottwitt, L. K. Oxenløwe, J. L. O'Brien, A. Laing, and M. G. Thompson, Science (2018).

[11] D. Bonneau, M. Lobino, P. Jiang, C. M. Natarajan, M. G. Tanner, R. H. Hadfield, S. N. Dorenbos, V. Zwiller, M. G. Thompson, and J. L. O'Brien, Phys. Rev. Lett. **108**, 053601 (2012).

[12] A. Laing, A. Peruzzo, A. Politi, M. R. Verde, M. Halder, T. C. Ralph, M. G. Thompson, and J. L. O'Brien, Appl. Phys. Lett. **97**, 211109 (2010).

[13] S. L. Braunstein and P. van Loock, Rev. Mod. Phys. **77**, 513 (2005).

[14] P. Jouguet, S. Kunz-Jacques, A. Leverrier, P. Grangier, and E. Diamanti, Nat. Photonics **7**, 378 (2013).

[15] The LIGO Scientific Collaboration, Nat. Photonics **7**, 613 (2013).

[16] T. Aoki, G. Takahashi, T. Kajiya, J. Yoshikawa, S. L. Braunstein, P. van Loock, and A. Furusawa, Nat. Phys. **5**, 541 (2009).

[17] S. Takeda, T. Mizuta, M. Fuwa, P. van Loock, and A. Furusawa, Nature **500**, 315 (2013).

[18] M. E. Anderson, J. D. Bierlein, M. Beck, and M. G. Raymer, Opt. Lett. **20**, 620 (1995).

[19] G. Masada, K. Miyata, A. Politi, T. Hashimoto, J. L. O'Brien, and A. Furusawa, Nat. Photonics **9**, 316 (2015).

[20] F. Kaiser, B. Fedrici, A. Zavatta, V. D'Auria, and S. Tanzilli, Optica **3**, 362 (2016).

[21] K. Yoshino, T. Aoki, and A. Furusawa, Appl. Phys. Lett. **90**, 41111 (2007).

[22] F. Lenzini, B. Haylock, J. C. Loredo, R. A. Abrahão, N. A. Zakaria, S. Kasture, I. Sagnes, A. Lemaitre, H.-P. Phan, D. V. Dao, P. Senellart, M. P. Almeida, A. G. White, and M. Lobino, Laser Photon. Rev. **11**, 1600297 (2017).

[23] M. Chen, N. C. Menicucci, and O. Pfister, Phys. Rev. Lett. **112**, 120505 (2014).

[24] Y. Cai, J. Roslund, G. Ferrini, F. Arzani, X. Xu, C. Fabre, and N. Treps, Nat. Commun. **8**, 15645 (2017).

[25] N. C. Menicucci, Phys. Rev. A **83**, 062314 (2011).

[26] S. Yokoyama, R. Ukai, S. C. Armstrong, C. Sornphiphatphong, T. Kaji, S. Suzuki, J. Yoshikawa, H. Yonezawa, N. C. Menicucci, and A. Furusawa, Nat. Photonics **7**, 982 (2013).

[27] N. C. Menicucci, P. van Loock, M. Gu, C. Weedbrook, T. C. Ralph, and M. A. Nielsen, Phys. Rev. Lett. **97**, 110501 (2006).

[28] F. Lenzini, S. Kasture, B. Haylock, and M. Lobino, Opt. Express **23**, 1748 (2015).

[29] H. Kogelnik and R. Schmidt, IEEE J. Quantum Electron. **12**, 396 (1976).

[30] S. Bennetts, G. D. McDonald, K. S. Hardman, J. E. Debs, C. C. N. Kuhn, J. D. Close, and N. P. Robins, Opt. Express **22**, 10642 (2014).

[31] W. P. Bowen, R. Schnabel, P. K. Lam, and T. C. Ralph, Phys. Rev. Lett. **90**, 043601 (2003).

[32] M. Zhang, C. Wang, R. Cheng, A. Shams-Ansari, and M. Lončar, Optica **4**, 1536 (2017).

**METHODS**

**Fabrication of the chip**

Waveguides were fabricated with a 1.85 μm proton exchange depth followed by annealing for 8 hours at 328 °C, and reverse proton exchange for 10 hours at the same temperature. Inputs of the periodically poled waveguides were designed with a channel width of 2.5 μm to get nearly single-mode operation at 775 nm and inject efficiently the pump beam into the fundamental mode of the waveguides. Channel width at the beginning of the poling region was increased to 8 μm with a 7 mm adiabatic taper to work in non-critical condition for quasi-phase matching. After the poling region, the channel widths are decreased to 6 μm with a second adiabatic taper of 1.5 mm length to get single-mode operation at 1550 nm. S-bends were designed with a sinusoidal function and a minimum bend radius of 40 mm. Separation between waveguide centres at the input and the output of the device was set to 127 μm to match the standard pitch of fibre V-groove arrays. To prevent back-reflections into the waveguides, the output facet was polished at an 8° angle. Total length of the chip is 62 mm.

The poling pattern was generated by standard electric-field poling with a period Λ = 16.12 μm, and a 50:50 duty cycle. After poling and waveguide fabrication, aluminium electrodes are realized on a 200 nm thick $SiO_2$ buffer layer in order to prevent optical absorption from the metal. Aluminium thickness was 250 nm, while electrodes were patterned using electron-beam lithography and wet etching.

Directional couplers were designed with separation between waveguide centres of 11.3 μm for DC1, DC4, and DC5, and 10.6 μm for DC2, and DC3. The lengths of the directional couplers are 6.1 mm for DC1, and 3.5 mm for DC2, DC3, DC4, and DC5. 12 mm long electrodes act as phase shifters on the local oscillator arms. At a zero applied voltage the SRs of the reconfigurable couplers are 72% for DC1, 85% for DC4, and 75% for DC5. Isolation of the pump beams by DC2 and DC3 on the adjacent homodyne detectors channels was found equal to 20 dB.

To prevent photorefractive damage, the chip was bonded with a UV curing glue to a custom-made aluminium oven and heated at 125 °C. Two printed circuit boards with SMA connectors were mounted on the sides of the oven and wire-bonded to the electrodes to control the voltage applied to phase shifters and directional couplers.

### Propagation losses

Transmission of the waveguides at the signal wavelength was tested on the second and fifth waveguides, and at the pump wavelength on the two central inputs. Transmission of the device corrected for Fresnel losses was found equal to 61% at 1550 nm, and to 40% at 775 nm. From numerical calculation of the mode overlap between waveguides and single-mode fibres[12], we estimated 0.14 dB/cm propagation losses at the signal wavelength, and 0.55 dB/cm propagation losses at the pump wavelength. Propagation losses at the signal wavelength were not directly measurable from the central inputs, since 1550 nm beams are only weakly guided in the first tapered section of the periodically poled waveguides.

### Detection efficiencies

Estimation of the detection efficiencies $\eta_1, \eta_2$ from Eq. 1 take into account 0.14 dB/cm propagation losses calculated from the centre of the periodically poled waveguides, a 0.5% loss introduced by the first directional coupler, 20% (for waveguide 1) and 14% (for waveguide 2) losses introduced by the pump filters DC2 and DC3, 13% Fresnel losses at the output facet, a 99% QE, and a 17 dB shot-noise clearance measured for a 4 mW local oscillator power.

### Shot-noise levels

To evaluate the shot-noise levels we used a motorized optical chopper blocking periodically the power of the pump beams. For each data acquisition, the shot-noise variance was calculated on five time windows with 0.4 ms duration. Standard error in the evaluation of the shot-noise levels was estimated equal to ±0.025 dB and added to all the uncertainties reported in the paper.

### Driving voltage

The electrodes on the chip were driven by three dual channel arbitrary waveform generators. The generators were operated in burst mode with a common trigger generated by a photodiode at the output of the optical chopper. Phase shifters, DC5, and DC1 (for generation of CV entanglement), required voltages in the ±10 V range and were driven directly by the waveform generators. DC4, and DC1 (for generation and homodyne detection of squeezed vacuum), required voltages of ±18 V and ±16 V respectively, generated with two voltage amplifiers. Low-pass filters from DC to 1.9 MHz were used to suppress unwanted amplitude noise introduced by the driving voltage in the measured side-bands.

### Squeezing and anti-squeezing levels

Squeezing and anti-squeezing levels were evaluated by fitting each noise trace with the function

$$\langle \Delta^2 \hat{X} \rangle = \langle \Delta^2 \hat{X}^+ \rangle \cos^2(at + \phi) + \langle \Delta^2 \hat{X}^- \rangle \sin^2(at + \phi),$$

where $t$ is the acquisition time and $a, \phi$ are fitting parameters. Uncertainties reported in the paper are the standard errors in the evaluation of the coefficients calculated by the least-square fitting procedure.

### Inseparability criterion

Variance of summed and subtracted quadratures were calculated from the photocurrents $i_1, i_2$ measured from the two homodyne detectors as[17]

$$\langle \Delta^2(\hat{X}_1 \pm \hat{X}_2) \rangle = \langle \Delta^2 \left( \frac{i_1}{\sqrt{2\langle \Delta^2 \hat{X}_{SN1} \rangle}} \pm \frac{i_2}{\sqrt{2\langle \Delta^2 \hat{X}_{SN2} \rangle}} \right) \rangle,$$

where $\langle\Delta^2 \hat{X}_{SN1}\rangle$, $\langle\Delta^2 \hat{X}_{SN2}\rangle$ are the shot-noise levels of the two homodyne detectors. Noise variances $\langle\Delta^2(\hat{X}_1^+ \pm \hat{X}_2^+)\rangle$, and $\langle\Delta^2(\hat{X}_1^- \pm \hat{X}_2^-)\rangle$, were calculated by averaging 4 points centered around the squeezed and anti-squeezed quadrature positions $X_1^+, X_2^+$, and $X_1^-, X_2^-$. Standard error in the evaluation of the noise levels was estimated as

$$SE = \frac{\sigma_{X_{1,2}}}{\sqrt{4}},$$

where $\sigma_{X_{1,2}}$ is the standard deviation of the noise traces measured on each homodyne detector. Due to the finite scanning speed of the second local oscillator phase shifter, the quadratures $X_1^-, X_1^+$, used for the calculation of the inseparability criterion have an offset of -0.09 rad and -0.10 rad respectively from the squeezed and anti-squeezed quadrature positions determined when the pump beams are in phase. We point out that the two quadratures are orthogonal within an offset which is smaller than the error in the quadrature positions (±0.02 rad) determined by fitting the data of Fig. 2b. Thus the measured data still satisfies the inseparability criterion, which is generally valid for any set of orthogonal quadratures.

## Homodyne detectors

The homodyne detectors used in this experiment use two matched photodetectors with custom ordered photodiodes from Laser Components with efficiencies of >99% and dark currents of >20pA. Each photodetector is in a dual amplifier configuration. The first stage uses a DC coupled transimpedance amplifier to amplify the photocurrent using the op amp AD829 with a transimpedance gain of 3k. The signal is then split in two using a resistor network for AC and DC coupled channels. The sidebands containing the measurements are present in the AC signal and the DC signal is used to monitor the detector. The AC path is filtered with a passive high pass filter with a corner frequency of 100 kHz and then amplified with a gain of 20 using another AD829. The DC coupled signal is amplified and used for monitoring. The noise floor of the AC coupled signals from each detector in a homodyne detector is matched using the compensation capacitor on the transimpedance amplifier. The AC signals are matched in phase using cable lengths and then subtracted to get the homodyne signal. The 3dB bandwidth of the homodyne detectors was measured to be 21 MHz and with a local oscillator power of 4 mW, they achieved a dark noise clearance of 17 dB below shot noise.

## SHG cavity

The SHG cavity is a free-space bow-tie configuration utilising a PPKTP nonlinear crystal. The cavity consists of two high-reflectivity (HR) concave mirrors at 1550 nm with ROC = 50 mm and two plane mirrors. The first plane mirror, input coupler (IC), is a partially reflecting mirror at 1550 nm; the second is a HR steering mirror attached to a piezo actuator. The cavity is locked on resonance utilising PDH technique. All mirrors are anti-reflection (AR) coated at the SHG wavelength. The cavity forms a beam waist of radius of approximately 27 μm between the two concave mirrors, where a 15 mm long PPKTP crystal is aligned. This configuration maximizes the non-linear conversion as detailed by the Boyd-Kleinman theory. The PPKTP crystal is housed in an oven and temperature stabilised around the optimum phase-matching temperature of 40°C using a Peltier element. Both faces of the crystal are wedged and AR coated at both wavelengths to minimise an intra-cavity parasitic interference. 1 W of fundamental optical power is injected through IC into the cavity and converted into the SHG wavelength with 80% efficiency. The SHG field exits the system through one of the concave mirrors and is subsequently coupled into an optical fibre.